\documentclass[twocolumn,secnumarabic,amssymb, amsmath, nofootinbib,tightenlines,
nobibnotes, aps, prl,epsfig]{revtex4}
\usepackage{graphicx}
\usepackage{dcolumn}
\usepackage{bm}
\begin{document}
\preprint{APS/123-QED}
\title{Higher order approximations to the longitudinal structure function $F_{L}$ from the parametrization of $F_{2}$ based on the  Laplace
transformation }

\author{G.R.Boroun}%
 \email{grboroun@gmail.com; boroun@razi.ac.ir }
\author{B.Rezaei }
\altaffiliation{brezaei@razi.ac.ir}
\affiliation{Department of Physics, Razi University, Kermanshah
67149, Iran}
\date{\today}
\begin{abstract}
We describe the determination of the longitudinal structure
function $F_{L}$ at NLO and NNLO approximations, using Laplace
transform techniques, into the parametrization of $F_{2}(x,Q^{2})$
and its derivative with respect to $\ln{Q^{2}}$ at low values of
the Bjorken variable $x$. The obtained results are comparable with
others by considering the effect of the charm quark mass to the
longitudinal structure function, which leads to rescaling variable
for $n_{f}=4$. Numerical calculations and comparison with H1 data
demonstrate that the suggested method provides reliable
$F_{L}(x,Q^{2})$ at small $x$ in a wide range of $Q^{2}$ values
and can be applied as well in analyses of ultra-high energy
processes with cosmic neutrinos. The obtained longitudinal
structure functions with and without the LHeC simulated
uncertainties [CERN-ACC-Note-2020-0002, arXiv:2007.14491 [hep-ex]
(2020)]  are compared with the H1 Collaboration data
[Eur.Phys.J.C{\bf74}, 2814(2014) and Eur.Phys.J.C{\bf71}, 1579
(2011)] and with the results from CT18 [Phys.Rev.D{\bf103},
014013(2021)] parametrization model at NLO and NNLO approximations.\\

\end{abstract}
 \pacs{***}
\maketitle
\subsection{Introduction}

In recent years, many attempts have been made to better understand
the longitudinal structure function experimentally and
theoretically [1-8]. In perturbative quantum chromodynamics
(pQCD), the longitudinal structure function contains information
about the gluon distribution and strong interaction dynamics. Thus
a measurement of the longitudinal proton structure function
provide a unique test of parton dynamics and the consistency of
QCD to the gluon density. The longitudinal structure function can
be extracted from the inclusive cross section only in the region
of large inelasticity $y$. At HERA, the measurement of the
longitudinal structure function collected about $5.9$ and
$12.2~\mathrm{Pb}^{-1}$ of data at reduced beam energies which
data were analysed together with about $100~\mathrm{Pb}^{-1}$ at
nominal HERA energies [9]. In ultra-high energy processes, at
extremely small $x$, the longitudinal structure function becomes
predominant and its  behavior  will be checked in high energy
process such as the Large Hadron electron Collider (LHeC)and the
Future Circular Collider electron-hadron (FCC-eh) projects which
runs to beyond a TeV in center-of-mass energy [9]. In the future
the electron-proton colliders  will be generated and extended much
lower values of $x$ and high values of $Q^{2}$. The simultaneous
measurement of the longitudinal structure functions  is the
cleanest way to establish new gluon density at small $x$. An
important advantage of future colliders, compared to HERA
experiments, is the wide range of $y$ values covered until 0.9.
Indeed the longitudinal structure function measurement will cover
an $x$ range from $2{\times}10^{-6}$ to above $x=0.01$ which the
LHeC promises to provide. As it extends the kinematic range in
electron-proton (ep) scattering by nearly four orders of magnitude
of ep collisions at HERA [10]. The interest in a measurement of
the longitudinal structure function, especially at small $x$, is
related to the uncertainty in the determination of the gluon
distribution. In this paper we deduce the longitudinal structure
function directly to the proton structure function uncertainty.\\
The longitudinal structure function is directly related to the
singlet and gluon distributions in the proton and its behavior
have been predicted by Altarelli and Martinelli [11] equation.
Authors in Ref.[11] derived an elegant formula for the
longitudinal structure function $F_{L}(x,Q^{2})$, also an effect
of order $\alpha_{s}(Q^{2})$, as a convolution integral over
$F_{2}(x,Q^{2})$ and the gluon density $g(x,Q^{2})$ by the
following form
\begin{eqnarray}
F_{L}(x,Q^{2})&=&C_{L,ns+s}(a_{s}(Q^{2}),x){\otimes}F_{2}(x,Q^{2})\nonumber\\
&&+<e^{2}>C_{L,g}(a_{s}(Q^{2}),x){\otimes}G(x,Q^{2}),
\end{eqnarray}
where $a_{s}(Q^{2})=\frac{\alpha_{s}(Q^{2})}{4\pi}$ and the
non-singlet densities become negligibly small in comparison with
the singlet densities at small $x$. Here $G(x,Q^{2})=xg(x,Q^{2})$
represents the gluon distribution function, $<e^{2}>$ is the
average of the charge $e^{2}$ for the active quark flavors,
$<e^{2}>=n_{f}^{-1}\sum_{i=1}^{n_{f}}e_{i}^{2}$ and the symbol
$\otimes$ denotes convolution according to the usual prescription.
$C_{L,i}(\mathrm{i=s,ns,g})$$^{,}$s are the coefficient function
which  can be written by the  perturbative expansion as follows
[12]
$$ C_{L,i} ^{(\varphi)} (a_{s},x)=\sum_{\phi=0}^{\varphi}a_{s}^{\phi+1}(Q^{2})c_{L,i}^{(\phi)}(x)
$$
where $\phi$ denotes the order in running coupling
$\alpha_{s}(Q^{2})$.\\
According to the DGLAP $Q^{2}$-evolution equations, the
 singlet distribution function leads to the following relation of integro-differential equation
\begin{eqnarray}
\frac{{\partial}F_{2}(x,Q^{2})}{{\partial}{\ln}Q^{2}}&=&-\frac{a_{s}(Q^{2})}{2}[P_{qq}(x){\otimes}F_{2}(x,Q^{2})\nonumber\\
&&+<e^{2}>P_{qg}(x){\otimes}xg(x,Q^{2})],\nonumber\\
\end{eqnarray}
where
\begin{eqnarray}
P_{a,b}(x)=P_{a,b}^{(0)}(x)
+a_{s}(Q^{2})\widetilde{P}_{a,b}^{(1)}(x)+a_{s}^{2}(Q^{2})\widetilde{P}_{a,b}^{(2)}(x)
\end{eqnarray}
and
\begin{eqnarray}
\widetilde{P}_{ab}^{(n)}(x)={P}_{ab}^{(n)}(x)+[C_{2,s}+C_{2,g}+...]\otimes
{P}_{ab}^{(0)}(x)+... .\nonumber
\end{eqnarray}
The quantities $\widetilde{P}_{ab}$$^{,}s$ are expressed via the
known splitting  and Wilson coefficient functions  in literatures
[13,14].\\
Recently authors in Ref.[4] revives the parametrization of the
longitudinal structure function at next-to-leading order (NLO)
approximation by using the parametrization of the proton structure
function $F_{2}(x,Q^{2})$ where suggested by authors in Ref.[15]
by a fit to HERA data on deep-inelastic lepton-nucleon scattering
(DIS) at small $x$. The parametrization of $F_{2}(x,Q^{2})$ is
relevant in investigations of ultra-high energy processes. Indeed
authors in Ref.[4] have obtained an analytical relation for the
longitudinal structure function at NLO approximation with respect
to the Mellin transform method, by the following form
\begin{eqnarray}
F_{L}^{\mathrm{NLO}}(x,Q^{2})&=&\tau(a_{s})\{
\vartheta(a_{s})F_{L}^{\mathrm{LO}}(x,Q^{2})\nonumber\\
&&-\chi(a_{s}^{2})F_{2}(x,Q^{2})\},
\end{eqnarray}
where
\begin{eqnarray}
F_{L}^{\mathrm{LO}}(x,Q^{2})& =&(1-
x)^{n}\sum_{\varepsilon=0}^{2}C_{\varepsilon}(Q^{2})L^{\varepsilon}\nonumber\\
\tau(a_{s})& =&[1+\frac{1}{3}a_{s}(Q^{2})L_{C}
(\widehat{\delta}^{(1)}_{sg}-\widehat{R}^{(1)}_{L,g})]^{-1}\nonumber\\
\vartheta(a_{s})&=&[1-a_{s}(Q^{2})
(\overline{\delta}^{(1)}_{sg}-\overline{R}^{(1)}_{L,g})]\nonumber\\
\chi(a_{s}^{2})&=&
a^{2}_{s}(Q^{2})[\frac{1}{3}\widehat{B}^{(1)}_{L,s}L_{A}+\overline{B}^{(1)}_{L,s}],
\end{eqnarray}
and
\begin{eqnarray}
F_{ 2}(x,Q^{2})& =&
D(Q^{2})(1-x)^{n}\sum_{\varepsilon=0}^{2}A_{\varepsilon}(Q^{2})L^{\varepsilon}.
\end{eqnarray}
Here $L^{,}$s are the logarithmic terms. The coefficient functions
at LO and NLO approximations are summarized in Appendix
A and  the effective parameters are defined in Table I.\\
In this article we investigate the behavior of the longitudinal
structure function inside the proton at high-order corrections to
the running coupling by using the Laplace transform techniques at
small $x$. Indeed, we use the Laplace-transform technique for
solving the Altarelli- Martinelli equation by employing the
parametrization of $F_{2}(x,Q^{2})$ at next-to-leading order (NLO)
and next-to-next-to-leading order (NNLO) aproximations. We
demonstrate that the small $x$ behavior of longitudinal structure
function can be directly related to the known structure function
$F_{2}(x,Q^{2})$ (i.e., Eq.(6)) and known its derivative
$\partial{F_{2}(x,Q^{2})}/\partial{\ln}{Q^{2}}$ at the higher order approximations.\\

\subsection{Method}
Considering the variable definitions $\upsilon{\equiv}\ln(1/x)$
and $w{\equiv}\ln(1/z)$, one can rewrite the equations (1) and (2)
in terms of the convolution integrals and new variables, as
\begin{eqnarray}
\frac{\partial{\mathcal{\widehat{F}}_{2}(\upsilon,Q^{2})}}{\partial{\ln}Q^{2}}&=&\int_{0}^{\upsilon}[\mathcal{\widehat{F}}_{2}(\upsilon,Q^{2})
\mathcal{\widehat{H}}^{(\varphi)}_{2,s}(a_{s}(Q^{2}),\upsilon-w)\\
&&+<e^{2}>\mathcal{\widehat{G}}(\upsilon,Q^{2})
\mathcal{\widehat{H}}^{(\varphi)}_{2,g}(a_{s}(Q^{2}),\upsilon-w)]dw,\nonumber\\
\mathcal{\widehat{F}}_{L}(\upsilon,Q^{2})&=&\int_{0}^{\upsilon}[\mathcal{\widehat{F}}_{2}(\upsilon,Q^{2})
\mathcal{\widehat{K}}^{(\varphi)}_{L,s+ns}(a_{s}(Q^{2}),\upsilon-w)\\
&&+<e^{2}>\mathcal{\widehat{G}}(\upsilon,Q^{2})
\mathcal{\widehat{K}}^{(\varphi)}_{L,g}(a_{s}(Q^{2}),\upsilon-w)]dw\nonumber,
\end{eqnarray}
where
\begin{eqnarray}
\mathcal{\widehat{F}}_{L}(\upsilon,Q^{2})&{\equiv}&F_{L}(e^{-\upsilon},Q^{2}),\nonumber\\
\frac{\partial{\mathcal{\widehat{F}}_{2}(\upsilon,Q^{2})}}{\partial{\ln}Q^{2}}&{\equiv}&
\frac{{\partial}F_{2}(e^{-\upsilon},Q^{2})}{\partial{\ln}Q^{2}},\nonumber\\
\mathcal{\widehat{G}}(\upsilon,Q^{2})&{\equiv}&G(e^{-\upsilon},Q^{2}),\nonumber\\
\mathcal{\widehat{H}}^{(\varphi)}(a_{s}(Q^{2}),\upsilon)&{\equiv}&e^{-\upsilon}\widehat{P}_{a,b}^{(\varphi)}(a_{s}(Q^{2}),\upsilon),\nonumber\\
\mathcal{\widehat{K}}^{(\varphi)}(a_{s}(Q^{2}),\upsilon)&{\equiv}&e^{-\upsilon}\widehat{C}_{L,i}^{(\varphi)}(a_{s}(Q^{2}),\upsilon).\nonumber
\end{eqnarray}
The Laplace transform of
$\mathcal{\widehat{H}}(a_{s}(Q^{2}),\upsilon)$$^{,}s$
 and $\mathcal{\widehat{K}}(a_{s}(Q^{2}),\upsilon)$$^{,}s$ are given by the following
 forms
\begin{eqnarray}
\Phi_{f}^{(\varphi)}(a_{s}(Q^{2}),s)&{\equiv}&
{\mathcal{L}}[\mathcal{\widehat{H}}^{(\varphi)}_{2,s}(a_{s}(Q^{2}),\upsilon);s]\nonumber\\
&&=\int_{0}^{\infty}\mathcal{\widehat{H}}^{(\varphi)}_{2,s}(a_{s}(Q^{2}),\upsilon)e^{-s\upsilon}d\upsilon,\nonumber\\
 \Theta_{f}^{(\varphi)}(a_{s}(Q^{2}),s)&{\equiv}&{\mathcal{L}}[\mathcal{\widehat{H}}^{(\varphi)}_{2,g}(a_{s}(Q^{2}),\upsilon);s]\nonumber\\
 &&=\int_{0}^{\infty}\mathcal{\widehat{H}}^{(\varphi)}_{2,g}(a_{s}(Q^{2}),\upsilon)e^{-s\upsilon}d\upsilon,\nonumber\\
\Phi_{L}^{(\varphi)}(a_{s}(Q^{2}),s)&{\equiv}&{\mathcal{L}}[\mathcal{\widehat{K}}^{(\varphi)}_{L,s+ns}(a_{s}(Q^{2}),\upsilon);s]\nonumber\\
&&=\int_{0}^{\infty}\mathcal{\widehat{K}}^{(\varphi)}_{L,s+ns}(a_{s}(Q^{2}),\upsilon)e^{-s\upsilon}d\upsilon,\nonumber\\
\Theta_{L}^{(\varphi)}(a_{s}(Q^{2}),s)&{\equiv}&{\mathcal{L}}[\mathcal{\widehat{K}}^{(\varphi)}_{L,g}(a_{s}(Q^{2}),\upsilon);s]\nonumber\\
&&=\int_{0}^{\infty}\mathcal{\widehat{K}}^{(\varphi)}_{L,g}(a_{s}(Q^{2}),\upsilon)e^{-s\upsilon}d\upsilon,\nonumber
\end{eqnarray}
with the condition $\mathcal{\widehat{H}}(\upsilon)=0$
 and $\mathcal{\widehat{K}}(\upsilon)=0$ for $\upsilon=0$ [16].
The convolution theorem for Laplace transforms allows us to
rewrite the right hand sides of Eqs.(7) and (8) with considering
the fact that the Laplace transform of the convolution factors are
simply the ordinary product of the Laplace transform of the
factors. Consequently, we can  obtain the equations for the
structure functions in the Laplace space $s$ by the following
forms as
\begin{eqnarray}
\frac{\partial{f_{2}(s,Q^{2})}}{\partial{\ln}Q^{2}}&=&
\Phi_{f}^{(\varphi)}(a_{s}(Q^{2}),s)f_{2}(s,Q^{2})\nonumber\\
&&+<e^{2}>\Theta_{f}^{(\varphi)}(a_{s}(Q^{2}),s)g(s,Q^{2}),\nonumber\\
f_{L}(s,Q^{2})&=&\Phi_{L}^{(\varphi)}(a_{s}(Q^{2}),s)f_{2}(s,Q^{2})\nonumber\\
&&+<e^{2}>\Theta_{L}^{(\varphi)}(a_{s}(Q^{2}),s)g(s,Q^{2}),
\end{eqnarray}
where
\begin{eqnarray}
{\mathcal{L}}[\mathcal{\widehat{F}}_{L}(\upsilon,Q^{2});s]&=&f_{L}(s,Q^{2}),\nonumber\\
{\mathcal{L}}[\mathcal{\widehat{F}}_{2}(\upsilon,Q^{2});s]&=&f_{2}(s,Q^{2}),\nonumber
\end{eqnarray}
and
\begin{eqnarray}
\eta_{j}^{(\varphi)}(a_{s}(Q^{2}),s)&=&\sum_{\phi=0}^{\varphi}a_{s}^{\phi+1}
(Q^{2})\eta^{(\phi)}_{j}(s),\nonumber\\
 &&\eta=(\Phi, \Theta
),~j=(f,L),\nonumber
\end{eqnarray}
where the superscript of the kernels represents the order in
$\alpha_{s}$. The leading-order coefficient functions $\Phi$ and
$\Theta$ in the Laplace space $s$ are given by
\begin{eqnarray}
\Phi_{L}^{(0)}(s)&=&4C_{F}\frac{1}{2+s},\nonumber\\
\Theta_{L}^{(0)}(s)&=&8n_{f}(\frac{1}{2+s}-\frac{1}{3+s}),\nonumber\\
\Theta_{f}^{(0)}(s)&=&2n_{f}(\frac{1}{1+s}-\frac{2}{2+s}+\frac{2}{3+s}),\\
\Phi_{f}^{(0)}(s)&=&4-\frac{8}{3}(\frac{1}{1+s}+\frac{1}{2+s}+2(\psi(s+1)+\gamma_{E})),\nonumber
\end{eqnarray}
where $\psi(x)$ is the digamma function and $\gamma_{E}=0.5772156
. . .$ is Euler constant.\\
Defining $\psi(s+1)+\gamma_{E}=S_{1}(s)$ and using the notion of
the so-called nested sums [4,17]. Let us to study the well known
function
$$
S_{a}(s)=\sum_{m=1}^{s}\frac{1}{m^{a}},
$$
where for the case $a{\geq}2$ is defined by the following form
\begin{eqnarray}
S_{a}(s)&=&[\sum_{m=1}^{\infty}-\sum_{m=s}^{\infty}]\frac{1}{m^{a}}=
S_{a}(\infty)-\sum\frac{1}{(l+s+1)^{a}}\nonumber\\
&&{\equiv}S_{a}(\infty)-\Psi_{a}(s+1).
\end{eqnarray}
Here $S_{a}(\infty)=\zeta(a)$ where $\zeta(a)$ is the Riemann zeta
function and
$\Psi_{a}(s+1)=\frac{(-1)^{a}}{(a-1)!}\Psi^{(a-1)}(s+1)$ where
$\Psi^{(a)}(s)$ is $a$-time derivative of the Euler
$\Psi$-function. Now let us to continue with the function
$$
S_{-a}(s)=\sum_{m=1}^{s}\frac{(-1)^{m}}{m^{a}},
$$
where by analogy with Eq.(11) we have that
\begin{eqnarray}
S_{-a}(s)&=&[\sum_{m=1}^{\infty}-\sum_{m=s}^{\infty}]\frac{(-1)^{m}}{m^{a}}\nonumber\\
&&=S_{-a}(\infty)-\sum\frac{(-1)^{l+s+1} }{(l+s+1)^{a}}\nonumber\\
&&{\equiv}S_{-a}(\infty)-(-1)^{s}\Psi_{-a}(s+1),
\end{eqnarray}
with $S_{-1}(\infty)={-\ln}2$ and
$S_{-a}(\infty)=\zeta(-a)=(2^{1-a}-1)\zeta(a)$ [17]. By analogy
with Eqs.(11) and (12), authors in Ref.[4] show that the functions
$S_{1}(s)$ and $S_{-1}(s)$ ones lead to the following functions
respectively
\begin{eqnarray}
S_{1}(s)&=&\Psi(s+1)-\Psi(1),\nonumber\\
S_{-1}(s)&=&-\ln(2)-\sum_{l=0}^{\infty}\frac{(-1)^{l+1}}{s+l+1}.
\end{eqnarray}
The above equation indicates that  for large $l^{,}$s , the
function is convergent, which is well known for any values of $s$.
In the following  we use the procedure of analytic continuation
for the sums $S_{1}(s)$, which come in consideration of the parton
distribution functions.\\
All further theoretical details relevant for analyzing $F_{L}$ at
NLO and NNLO in the $\mathrm{\overline{MS}}$ factorization scheme
have been presented in Refs.[18-21]. The explicit expressions for
the NLO and NNLO kernels in $s$-space are rather cumbersome,
therefore we recall that we are interested in investigation of the
kernels in small $x$ [12, 18-21]. In the Laplace space we consider
the kernels at  small $s$, as the two and three-loop kernels read
\begin{eqnarray}
\Phi_{L,s{\rightarrow}0}^{(1)}(s)&{\simeq}&n_{f}[-\frac{2.371}{s}],\nonumber\\
\Theta_{L,s{\rightarrow}0}^{(1)}(s)&{\simeq}&n_{f}[-\frac{5.333}{s}],\nonumber\\
\Theta_{f,s{\rightarrow}0}^{(1)}(s)&{\simeq}&C_{A}T_{f}[\frac{40}{9s}],\nonumber\\
\Phi_{f,s{\rightarrow}0}^{(1)}(s)&{\simeq}&C_{F}T_{f}[\frac{40}{9s}],
\end{eqnarray}
and
\begin{eqnarray}
\Phi_{L,s{\rightarrow}0}^{(2)}(s)&{\simeq}&n_{f}[-\frac{885.530}{s}+\frac{182}{s^{2}}]+n^{2}_{f}[\frac{40.239}{s}],\nonumber\\
\Theta_{L,s{\rightarrow}0}^{(2)}(s)&{\simeq}&n_{f}[-\frac{2044.700}{s}+\frac{409.506}{s^{2}}]+n^{2}_{f}[\frac{88.504}{s}],\nonumber\\
\Theta_{f,s{\rightarrow}0}^{(2)}(s)&{\simeq}&n_{f}[-\frac{1268.300}{s}+\frac{896}{3s^{2}}]+n^{2}_{f}[\frac{1112}{243s}],\nonumber\\
\Phi_{f,s{\rightarrow}0}^{(2)}(s)&{\simeq}&n_{f}[-\frac{506}{s}+\frac{3584}{27s^{2}}]+n^{2}_{f}[\frac{256}{81s}].
\end{eqnarray}
The standard representation for QCD couplings in NLO and NNLO
(within the $\mathrm{\overline{MS}}$-scheme) approximations have
the forms
\begin{eqnarray}
\alpha_{s}(t)&=&\frac{4\pi}{\beta_{0}t}\Big{[}1
-\frac{\beta_{1}}{\beta_{0}^{2}}\frac{\ln{t}}{t}\Big{]}~~~~~~~~~~~~~~~~~~~~~~~~~~~~~(\mathrm{NLO}),\nonumber\\
\alpha_{s}(t)&=&\frac{4\pi}{\beta_{0}t}\Big{[}1
-\frac{\beta_{1}}{\beta_{0}^{2}}\frac{\ln{t}}{t}\nonumber\\
&&+\frac{1}{\beta_{0}^{3}t^{2}}\bigg{\{}
\frac{\beta_{1}^{2}}{\beta_{0}}(\ln^{2}t-\ln{t}-1)+\beta_{2}\bigg{\}}
\Big{]}~~~(\mathrm{NNLO}),\nonumber
\end{eqnarray}
where $\beta_{0}$, $\beta_{1}$ and $\beta_{2}$ are the one, two
and three loop correction to the QCD $\beta$-function and
$t=\ln\frac{Q^{2}}{\Lambda^{2}}$, $\Lambda$ is the QCD cut-off
parameter.\\
Consequently, by working in the Laplace space $s$, we can obtain
the longitudinal structure function by solving Eq.(9) for
$f_{L}(s,Q^{2})$ into $f_{2}(s,Q^{2})$ and
${\partial{f_{2}(s,Q^{2})}}/{\partial{\ln}Q^{2}}$ as
\begin{eqnarray}
f_{L}(s,Q^{2})&=&k^{(\varphi)}(a_{s}(Q^{2}),s)f_{2}(s,Q^{2})\nonumber\\
&&+h^{(\varphi)}(a_{s}(Q^{2}),s)\frac{\partial{f_{2}(s,Q^{2})}}{\partial{\ln}Q^{2}},
\end{eqnarray}
where the kernels $k^{(\varphi)}(a_{s}(Q^{2}),s)$ and
$h^{(\varphi)}(a_{s}(Q^{2}),s)$ contain contributions of the
$s$-space splitting and coefficient functions up to the NNLO
approximation. These kernels can be evaluated from $s$-space
results by the following forms
\begin{eqnarray}
k^{(\varphi)}(a_{s}(Q^{2}),s)&=&\sum_{\phi=0}^{\varphi}
a_{s}^{\phi+1}(Q^{2})\Phi^{(\phi)}_{L}(s)\nonumber\\
&&-h^{(\varphi)}(a_{s}(Q^{2}),s)\sum_{\phi=0}^{\varphi}a_{s}^{\phi+1}(Q^{2})\Phi^{(\phi)}_{f}(s),\nonumber\\
h^{(\varphi)}(a_{s}(Q^{2}),s)&=&\frac{\sum_{\phi=0}^{\varphi}a_{s}^{\phi+1}(Q^{2})
\Theta^{(\phi)}_{L}(s)}{\sum_{\phi=0}^{\varphi}a_{s}^{\phi+1}(Q^{2})\Theta^{(\phi)}_{f}(s)}.
\end{eqnarray}
 The inverse Laplace transform of coefficients $k(a_{s}(Q^{2}),s)$ and $h(a_{s}(Q^{2}),s)$ in above equations are
defined as kernels
$$\widehat{\eta}(a_{s}(Q^{2}),\upsilon){\equiv}{\mathcal{L}}^{-1}[k(a_{s}(Q^{2}),s);\upsilon]$$
and
$$\widehat{J}(a_{s}(Q^{2}),\upsilon){\equiv}{\mathcal{L}}^{-1}[h(a_{s}(Q^{2}),s);\upsilon]$$
respectively. Clearly the kernels (i.e., $\widehat{\eta}$ and
$\widehat{J}$) are dependent on $\upsilon$ and the running
coupling at the higher order approximations. We will generally not
be able to define an analytical form for these kernels at higher
order approximations, so $F_{L}$ determined by  numerical integral
of the parametrization of $F_{2}$ and its derivative, as
\begin{eqnarray}
\widehat{F}_{L}(\upsilon,Q^{2})&{\equiv}&{\mathcal{L}}^{-1}[f_{L}(s,Q^{2});\upsilon]\\
&&=\int_{0}^{\upsilon}[\widehat{F}_{2}(w,Q^{2})\widehat{\eta}^{(\varphi)}(a_{s}(Q^{2}),\upsilon-w)\nonumber\\
&&+\frac{\partial{\widehat{F}_{2}(w,Q^{2})}}{\partial{\ln}Q^{2}}\widehat{J}^{(\varphi)}(a_{s}(Q^{2}),\upsilon-w)]dw.\nonumber
\end{eqnarray}
Consequently, one can obtain the longitudinal structure function
as $F_{L}(x,Q^{2})$. Therefore the general analytical expression
for the longitudinal structure function in $x$-space is given by
\begin{eqnarray}
F_{L}(x,Q^{2})&=&\int_{x}^{1}\frac{dy}{y}[F_{2}(y,Q^{2})\eta^{(\varphi)}({\frac{x}{y}},Q^{2})\nonumber\\
&&+\frac{{\partial}F_{2}(y,Q^{2})}{\partial{\ln}Q^{2}}J^{(\varphi)}({\frac{x}{y}},Q^{2})].
\end{eqnarray}
So that we have an explicit solution for the longitudinal
structure function at NLO and NNLO approximations which can be
evaluated to the numerical accuracy to which $F_{2}(x,Q^{2})$ is
known. Having an analytical proton structure function and its
derivative with respect to $\ln{Q^{2}}$, one can extract the
longitudinal structure function numerically at any desired $x$ and
$Q^{2}$ values.\\

\subsection{Results and Discussion}

In order to make the effect of production threshold for charm
quark at $n_{f}=4$ one should take into account quark mass for
small $Q^{2}$. To this end we shall follow the rescaling variable
$\chi$ which introduced by Aivazis, Collins, Olness and Tung
(ACOT) in Ref.[22]. Therefore, the longitudinal structure function
is defined by the rescaling variable $\chi$ where
$$ \chi=x(1+\frac{4m_{c}^{2}}{Q^{2}}).
$$
This rescaled variable is one of the ingredients used in the
general-mass variable flavor number scheme (GM-VFNS), which is
used in the global fits of PDFs of the CETQ and MRST groups. The
running charm mass is obtained as $m_{c}=1.29^{+0.077}_{-0.053}
\mathrm{GeV}$, where the uncertainties are obtained through adding
the experimental fit, model and parametrization uncertainties in
quadrature [1,2]. At high $Q^{2}$ values
($m_{c}^{2}/Q^{2}{\ll}1$), the rescaling variable $\chi$ reduces
to the Bjorken variable $x$ as $\chi{\rightarrow}x$ [22,23].  The
QCD parameter $\Lambda$ has been extracted due to
$\alpha_{s}(M_{z}^{2})=0.1166$, which for four number of active
flavor is defined by $\Lambda_{\mathrm{QCD}}^{\mathrm{LO}}=136.8~
\mathrm{MeV}$ and $\Lambda_{\mathrm{QCD}}^{\mathrm{NLO}}=284.0~
\mathrm{MeV}$. Also we take
$\Lambda_{\mathrm{QCD}}^{\mathrm{NNLO}}=235.0~ \mathrm{MeV}$.\\
Now we can proceed to extract the longitudinal structure function
$F_{L}(x,Q^{2})$ with the explicit form of the proton structure
function and its derivative at NLO and NNLO approximations. In
order to present more detailed discussions on our findings, the
results for the longitudinal structure function compared with CT18
[24] parametrization model. It should also be mentioned that CT18
results at NLO and NNLO approximations obtained using a wide
variety of high-precision Large Hadron Collider (LHC) data, in
addition to the combined HERA I+II deep inelastic scattering data
set. In Fig.1 we are presented the $x$-dependence of the
longitudinal structure function at $Q^{2}=5, 15, 25$ and
$45~\mathrm{GeV}^{2}$ and compared with H1 Collaboration data
[1,2] and the results from CT18 NLO parametrization  model. The
error bands illustrated in this figure, and in the other figures,
are into the charm-quark mass uncertainty and the statistical
errors in the parametrization of $F_{2}(x,Q^{2})$ and its
derivative, where the fit parameter errors are shown in Table I.
As can be seen from the related figures, the longitudinal
structure function results are consistent with the CT18 NLO and H1
Collaboration data at moderate and large values of $Q^{2}$. It is
seen that, for all values of the presented $Q^{2}$ with respect to
the rescaling variable, the extracted longitudinal structure
functions at NLO approximation due to the Laplace transform method
are in a good agreement with data and parametrization models. \\
In Fig.2, the results for the longitudinal structure function
within the NNLO approximation have been shown and compared with
the NNLO analysis of CT18 model. We observe that, with respect to
the approximation approach used in the coefficient functions
 at higher order approximation in the limit $x{\rightarrow}0$, the extracted longitudinal structure functions within the NNLO
approximation are comparable with the experimental data and the
CT18 NNLO model. These results are interesting in connection with
theoretical investigations of ultra-high energy processes with
cosmic neutrinos.\\
In Fig.3 the longitudinal structure function results at NLO and
NNLO approximations due to the Laplace transforms method are
associated with the LHeC simulated uncertainties. These simulated
uncertainties for the longitudinal structure function measurement
recently reported by the LHeC Collaboration and FCC-he Study Group
in Ref.[9]. In this figure the straight lines represent the CT18
NLO and CT18 NNLO QCD analysis in different schemes and the up and
down triangles represent our results as accompanied with the LHeC
simulated uncertainties. We compare the results for the
longitudinal structure function at NLO and NNLO approximations
with a general-mass variable-flavor-number scheme (GM-VFNS) and
zero-mass variable-flavor-number scheme (ZM-VFNS) in the CT18 NLO
and NNLO methods  in this figure respectively. As can be seen from
the related figures, the longitudinal structure function results
are consistent with different schemes in the CT18 NLO and NNLO at
moderate and large values of $Q^{2}$.\\
In Fig.4, we show the $Q^{2}$- dependence of the longitudinal
structure function at small $x$ at NLO approximation. In this
figure (i.e., Fig.4) the results of calculations and comparison
with the H1 collaboration data [1,2] are presented. These results
have been performed at fixed value of the invariant mass $W$ as
$W=230~ \mathrm{GeV}$. Over a wide range of the variable $Q^{2}$,
the extracted longitudinal structure functions at NLO
approximation are in a good agreement with experimental data and
CT18 NLO analysis. For $Q^{2}<1~\mathrm{GeV}^{2}$, the extracted
results have the same CT18 NLO  behavior, but there are no data to
compare in this region. The error bands illustrated in this figure
are into the charm-quark mass uncertainty and the statistical
errors in the parametrization of $F_{2}(x,Q^{2})$, where the fit
parameter errors are shown in Table I.\\
The longitudinal structure function behavior with the NNLO
approximation is shown in Fig.5 for a wide range of $Q^{2}$. These
results in Fig.5 compared with the H1 collaboration data [1,2] and
CT18 NNLO. As can be seen in this figure, these results based on
the Laplace transforms method  are comparable with the CT18 NNLO
analysis. However, at extremely low momenta,
$Q^{2}<1~\mathrm{GeV}^{2}$, the extracted $F_{L}$ within NNLO
approximation  is
 below the experimental data. In this region
the depletion and enhancement of $F_{L}$ have the same behavior in
comparison with the CT18 NNLO model. One wishes to improve
substantially the precision of the $F_{L}$ data with extension of
kinematic range at the LHeC and FCC-eh for testing theory at small
$x$ and small $Q^{2}$ values. As commented in Refs.[9,25],
resummation of the large $\ln(1/x)$ terms restore the  dominance
of the $gg$ splitting over the $qg$ one. The resummation of
$(\alpha_{s}\ln{s})^{n}$ series in the leading logarithmic order
is a differential equation in $\ln(1/x)$ for small $x$ evolution
equation. The leading logarithmic (LLx) results yielded a growth
of the gluon density and the next-to-leading logarithmic (NLLx)
calculation yielded  some instability in the cross sections. The
appearance of the large negative corrections at NLLx motivated the
longitudinal structure functions for the appropriate resummation
which would stabilize the results. It was demonstrated that the
resummed fits provide a better description of the longitudinal
structure function data than the pure method based fits at fixed
NNLO approximation. Such effects will be strongly magnified at the
LHeC, as it was shown that the description of the longitudinal
structure function from HERA data is improved in the fits with the
small $x$ resummation. This analysis suggests  that the small $x$
resummation effects will be visible in the small $x$ and small
$Q^{2}$ region. Indeed the longitudinal structure function in this
region at NNLO approximation increase due to the resummation
predictions
 as $x$ decreases.\\
 In Fig.6 the longitudinal structure function results at NLO and
 NNLO approximations are compared with the Regge-like behavior of
 the parton distribution functions. In Ref.[26] authors extracted
 a formula for the longitudinal structure function $F_{L}$ as
 function of $F_{2}$ and its derivative at small $x$ at LO and NLO
 approximations based on the Regge-like behavior. The Regge-like behavior for the
 singlet and gluon distribution functions at small $x$ is given by
 $$
G(x,Q^{2})=x^{-\delta}\widetilde{G}(x,Q^{2}),
~~F_{2}(x,Q^{2})=x^{-\delta}\widetilde{s}(x,Q^{2})
 $$
where the $\delta $ value obtained by fixed coupling LLx BFKL
gives $\delta{\simeq}0.5$, which is the so-called hard-Pomeron
exponent. This value was obtained in the studies performed in Ref.
[27] as the sum of the leading powers of $\ln(1/x)$ in all orders
of perturbation theory. In tensor-Pomeron model [28], where in
addition to the soft tensor Pomeron a hard tensor Pomeron and
Reggeon exchange included, the hard-Pomeron intercept was
determined to be $\delta{\simeq}0.3$. In Fig.6 our results
compared with the longitudinal structure function extracted in
Ref.[26] based on the parameterization of $F_{2}$ (i.e., Eq.6). We
compared these results with the  H1 collaboration data [1,2] and
the Regge-like behavior in Ref.[26] at
$Q^{2}=20~\mathrm{GeV}^{2}$. The results at NLO and NNLO
approximations are comparable with the H1 collaboration data.\\
Finally we analysis the coefficient functions at NLO and NNLO
approximations for the behavior of the longitudinal structure
functions at small $x$ in Fig.7. Authors in Refs.[18,29]
considered the dynamical and standard distributions at small $x$.
The gluon distribution at dynamical has a steeper behavior at
small $x$ in comparison with the standard model. Also the sea
distribution has a similar behavior at the standard and dynamical
models. Authors shown that at NLO approximation the longitudinal
structure function in dynamically distribution is larger than the
NNLO one for $Q^{2}{\geq}5~\mathrm{GeV}^{2}$ at small $x$.  A
similar behavior prevails for the longitudinal structure function
in Fig.3 at moderate and large $Q^{2}$ due to the results of CT18
NLO and NNLO approximations in the GM-VFNS and ZM-VFNS. The
coefficient functions in (14) and (15) are shown in Fig.7 at
$s$-space. The behavior of the coefficient functions is considered
at NLO and NNLO approximations for $s{\geq}0$. It is evidence from
these behaviors in Fig.7 that  at NNLO the longitudinal structure
function values are less then the $F_{L}$-values at NLO. Also the
leading twist-2 predictions are necessary for illustrate of the
longitudinal structure function behavior at $Q^{2}<5~\mathrm{GeV}^{2}$.\\
In conclusion, we have presented a certain theoretical model at
NLO and NNLO approximations to describe the longitudinal structure
function  based on the Laplace transforms method  at small values
of $x$. Indeed, there are various methods to solve the
Altarelli-Martinelli equation to obtain the longitudinal structure
function, in this manuscript we have shown that the method of the
Laplace transforms technique is also the reliable and alternative
scheme to solve Altarelli-Martinelli equation, analytically. A
detailed analysis has been performed to find an analytical
solution of the longitudinal structure function into the
parametrization of $F_{2}(x,Q^{2})$ and its derivative of the
proton structure function with respect to ${\ln}Q^{2}$ at
high-order corrections. The calculations are consistent with the
H1 data from HERA collider and they are comparable with the CT18
at NLO and NNLO approximations. Also we compared the longitudinal
structure functions with respect to the LHeC simulated
uncertainties with the CT18 at NLO and NNLO approximations due to
the GM-VFN and ZM-VFN schemes. This persuades us that the obtained
results can be extended to high energy regime in new colliders
(like in the proposed LHeC and FCC-eh colliders). These results
indicate that the obtained solutions from present analysis at NLO
and NNLO approximations based on Laplace transform technique are
comparable with the ones obtained by global QCD analysis of CT18
from the parton distribution functions. In all figures  clearly
demonstrate that the extraction procedure provides correct
behaviors of the extracted longitudinal structure function  in
both NLO and NNLO approximations. At intermediate and high $Q^{2}$
the extracted longitudinal structure functions at NLO and NNLO
approximations are in a good agreement with experimental data.
Indeed, for very small $Q^{2}$ values, the NNLO+NNLx resummation
will be improve the longitudinal structure function behavior at
NNLO approximation in the future colliders. We also showed that
the obtained results from the longitudinal structure function
analysis are in good
agreement with those from the literature.\\

\subsection{ACKNOWLEDGMENTS}

The authors are thankful to the Razi University for financial support of this project.
 Also G.R.Boroun thanks M.Klein and N.Armesto for allowing access
to data related to simulated errors of the longitudinal structure
function
at the Large Hadron electron Collider (LHeC).\\

\subsection{Appendix A}
The coefficient functions read as
\begin{eqnarray}
C_{2}&=&\widehat{A}_{2}+\frac{8}{3}a_{s}(Q^{2})DA_{2}\nonumber\\
C_{1}&=&\widehat{A}_{1}+\frac{1}{2}\widehat{A}_{2}+\frac{8}{3}a_{s}(Q^{2})D[A_{1}+(4\zeta_{2}-\frac{7}{2})A_{2}]\nonumber\\
C_{0}&=&\widehat{A}_{0}+\frac{1}{4}\widehat{A}_{2}-\frac{7}{8}\widehat{A}_{2}+\frac{8}{3}a_{s}(Q^{2})D[A_{0}
+(2\zeta_{2}-\frac{7}{4})A_{1}\nonumber\\
&&+(\zeta_{2}-4\zeta_{3}-\frac{17}{8})A_{2}],
\end{eqnarray}
\begin{eqnarray}
\widehat{A}_{2}&=&\widetilde{A}_{2}\nonumber\\
\widehat{A}_{1}&=&\widetilde{A}_{1}+2DA_{2}\frac{\mu^{2}}{\mu^{2}+Q^{2}}\nonumber\\
\widehat{A}_{0}&=&\widetilde{A}_{0}+DA_{1}\frac{\mu^{2}}{\mu^{2}+Q^{2}}\nonumber\\
\widetilde{A}_{i}&=&\widetilde{D}A_{i}+D\overline{A}_{i}\frac{Q^{2}}{Q^{2}+\mu^{2}}\nonumber\\
\widetilde{D}&=&\frac{M^{2}Q^{2}[(2-\lambda)Q^{2}+\lambda
M^{2}]}{[Q^{2}+M^{2}]^{3}}\nonumber\\
\overline{A}_{\varepsilon}&=&a_{\varepsilon1}+2a_{\varepsilon2}L_{2},~~a_{02}=0.
\end{eqnarray}
and
\begin{eqnarray}
\widehat{B}^{(1)}_{L,s}&=&8C_{F}[\frac{25}{9}n_{f}-\frac{449}{72}C_{F}+(2C_{F}-C_{A})\nonumber\\
&&(\zeta_{3}+2\zeta_{2}-\frac{59}{72})]\nonumber\\
\overline{B}^{(1)}_{L,s}&=&\frac{20}{3}C_{F}(3C_{A}-2n_{f})\nonumber\\
\widehat{\delta}^{(1)}_{sg}&=&\frac{26}{3}C_{A}\nonumber\\
\overline{\delta}^{(1)}_{sg}&=&3C_{F}-\frac{347}{18}C_{A}\nonumber\\
\widehat{R}^{(1)}_{L,g}&=&-\frac{4}{3}C_{A}\nonumber\\
\overline{R}^{(1)}_{L,g}&=&-5C_{F}-\frac{4}{9}C_{A}\nonumber\\
L_{A}&=&L+\frac{A_{1}}{2A_{2}}\nonumber\\
L_{C}&=&L+\frac{C_{1}}{2C_{2}}\nonumber\\
L&=&\ln(1/x)+L_{1}\nonumber\\
L_{1}&=&{\ln}\frac{Q^{2}}{Q^{2}+\mu^{2}}\nonumber\\
L_{2}&=&{\ln}\frac{Q^{2}+\mu^{2}}{\mu^{2}}\nonumber\\
A_{i}(Q^{2})&=&\sum_{k=0}^{2}a_{ik}L_{2}^{k},~ (i=1,2)\nonumber\\
A_{0}&=&a_{00}+a_{01}L_{2}\nonumber\\
D&=&\frac{Q^{2}(Q^{2}+\lambda M^{2})}{(Q^{2}+M^{2})^2},
\end{eqnarray}
with the color factors $C_{A}=3$ and $C_{F}=\frac{4}{3}$
associated with the color group $SU(3)$ and $n_{f}$ being the
number of flavors.\\
\begin{table}
\caption{ The effective parameters at small $x$ for
$0.15~\mathrm{GeV}^{2}<Q^{2}<3000~\mathrm{GeV}^{2}$ provided by
the following values. The fixed  parameters are defined by the
Block-Halzen fit to the real photon-proton cross section as
$M^{2}=0.753 \pm 0.068~ \mathrm{GeV}^{2}$, $\mu^2 = 2.82 \pm
0.290~ \mathrm{GeV}^{2}$, $n=11.49\pm 0.99$ and $\lambda=
2.430~\pm 0.153$  [15].}
\begin{tabular} {cccc}
\toprule \\  \multicolumn{2}{c}{parameters \quad \quad \quad ~~~~~~~~~~~~~~~~value}    \\ &&&\\ \hline \\ &&&\\
  $a_{10} $  &   \quad  $8.205\times 10^{-4}~~  \pm  4.62\times10^{-4} $  \\

  $a_{11} $  &   \quad   $-5.148\times 10^{-2}\pm 8.19\times10^{-3}$  \\

  $a_{12}$   &    \quad  $-4.725\times 10^{-3}\pm 1.01\times10^{-3}$   \\  &&&\\

 $a_{20}$   &   \quad   $2.217\times 10^{-3}\pm 1.42\times10^{-4} $ \\

 $a_{21}$   &   \quad   $1.244\times 10^{-2}\pm 8.56\times10^{-4}$  \\

 $a_{22}$    &    \quad  $5.958\times 10^{-4}\pm 2.32\times10^{-4} $ \\ &&& \\

$a_{00}$& \quad  $2.550\times 10^{-1}~\pm 1.600\times10^{-2}$ & &\\

$a_{01}$& \quad  $1.475\times 10^{-1}~\pm 3.025\times10^{-2}$ & &\\

\hline

\end{tabular}
\end{table}

\newpage{
\section{References}

1. H1 and ZEUS Collaborations (H. Abramowicz  et al.), Eur. Phys.
J. C {\bf78}, 473 (2018).\\
2. H1 Collab. (V.Andreev  et al.), Eur.Phys.J.C{\bf74},
2814(2014); H1 Collab. (F.D.Aaron  et al.),Eur.Phys.J.C71, 1579 (2011).\\
3. L.P.Kaptari et al., JETP Lett.{\bf 109}, 281(2019).\\
4. L.P.Kaptari et al., Phys.Rev.D {\bf99}, 096019 (2019).\\
5. V.Tvaskis et al., Phys.Rev.C {\bf97}, 045204 (2018).\\
6. G.R.Boroun, arXiv: 2108.09465 [hep-ph]; JETP Lett. {\bf114}, 1
(2021); Eur.Phys.J.Plus {\bf135}, 68 (2020); Phys.Rev.C {\bf97},
015206 (2018).\\
7. M.Niedziela and M.Praszalowicz, Acta Phys.Polon.B {\bf46}, 2019
(2015); N.Baruah, M.K.Das and J.K.Sarma, Eur.Phys.J.Plus {\bf129},
229 (2014); M.Mottaghizadeh and A.Mirjalili, Phys.Lett.B {\bf820},
136534 (2021);
A.D.Martin, W.J.Stirling and R.S.Thorne, Phys.Lett.B {\bf635}, 305 (2006);
A.D.Martin, W.J.Stirling and R.S.Thorne, Phys.Lett.B {\bf636}, 259 (2006); S.Zarrin and S.Dadfar, Int.J.Theor.phys.{\bf60}, 3822(2021).\\
8. B.Rezaei and G.R.Boroun, Eur.Phys.J.A{\bf56}, 262 (2020);
L.Ghasemzadeh, A.Mirjalili and S.Atashbar Tehrani, Phys.Rev.D
{\bf104}, 074007 (2021); G.R.Boroun and B.Rezaei, Chin.Phys.Lett.
{\bf32}, 111101 (2015); S.S.Mohsenabadi, S.Atashbar Tehrani and
F.Taghavi-Shahri, arXiv:2112.03373; G.R.Boroun, B.Rezaei and
J.K.Sarma, Int.J.Mod.Phys.A {\bf29}, 1450189 (2014); S.Shoeibi,
F.Taghavi-Shahri, H.Khanpour and K.Javidan, Phys.Rev.D {\bf97},
074013 (2018); G.R.Boroun and B.Rezaei, Eur.Phys.J.C {\bf72}, 2221
(2012); G.R.Boroun and B.Rezaei, Phys.Letts.B {\bf816}, 136274
(2021); H.Khanpour, A.Mirjalili and
S.Atashbar Tehrani, Phys.Rev.C {\bf95}, 035201 (2017).\\
9. LHeC Collaboration and FCC-he Study Group, P. Agostini et al.,
J. Phys. G: Nucl. Part. Phys. {\bf48}, 110501(2021).\\
10 M. Klein, arXiv: 1802.04317[hep -ph]; Ann. Phys.
{\bf528}, 138 (2016).\\
11. G.Altarelli and G.Martinelli, Phys.Lett.B\textbf{76}, 89(1978).\\
12. S. Moch, J.A.M. Vermaseren, and A. Vogt, Phys. Lett. B
{\bf606}, 123 (2005).\\
13. J. Blumlein, V. Ravindran and W. van Neerven, Nucl. Phys. B
\textbf{586}, 349(2000); S.Catani and F.Hautmann,
Nucl.Phys.B{\bf427}, 475(1994).\\
14. D.I.Kazakov and A.V.Kotikov, Phys.Lett.B{\bf291}, 171(1992);
E.B.Zijlstra and W.L.van Neerven, Nucl.Phys.B{383}, 525(1992).\\
15. M. M. Block, L. Durand and P. Ha, Phys. Rev. D {\bf89}, 094027
(2014).\\
16. M.M.Block, Eur.Phys.J.C {\bf65}, 1 (2010); M.M.Block, L.Durand
and D.W.McKay, Phys.Rev.D {\bf79}, 014031 (2009).\\
 17. A. V. Kotikov and V. N. Velizhanin, arXiv:
0501274 [hep-ph]
(2005); D.I.Kazakov and A.V.Kotikov, Phys.Lett.B {\bf291}, 171 (1992).\\
18. M.Gl$\mathrm{\ddot{u}}$k, C.Pisano and E.Reya,
Phys.Rev.D{\bf77}, 074002 (2008).\\
19. C.D.White and R.S.Thorne, Eur.Phys.J.C {\bf45}, 179 (2006).\\
20. A. Vogt, S. Moch and J.A.M. Vermaseren, Nucl.Phys.B {\bf691},
129 (2004).\\
21. W.L. van Neerven and A.Vogt, Phys.Lett.B {\bf490}, 111
(2000).\\
22. M.A.G.Aivazis et al., Phys.Rev.D {\bf50}, 3102 (1994).\\
23. A.V.Kotikov, B.G.Shaikhatdenov and Pengming Zhang,
Phys.Rev.D96, 114002(2017); G.Beuf, C.Royon and D.Salek, arXiv [hep-ph]:0810.5082.\\
24. T.-J. Hou, et al., Phys. Rev. D {\bf103}, 014013 (2021).\\
25. H. Abdolmaleki et al., Eur. Phys. J. C {\bf78}, 621 (2018).\\
26. A.V. Kotikov, JETP {\bf80}, 979 (1995); A.V. Kotikov, G.
Parente, Mod.Phys.Lett.A {\bf12}, 963 (1997).\\
27. E.A.Kuraev, L.N.Lipatov and V.S.Fadin, ZHETF {\bf53}, 2018
(1976); {\bf54}, 128 (1977); Ya.Ya.Balitzki and L.N.Lipatov,
Yad.Fiz. {\bf28}, 822 (1978); L.N.Lipatov, ZHETF {\bf63},904
(1986).\\
28. D.Britzger et al., Phys. Rev. D {\bf100}, 114007 (2019).\\
29. C.Pisano, Nuclear Physics B-Proceedings Supplements,
{\bf186}, 47 (2009).\\
 }
\begin{figure}
\includegraphics[width=1\textwidth]{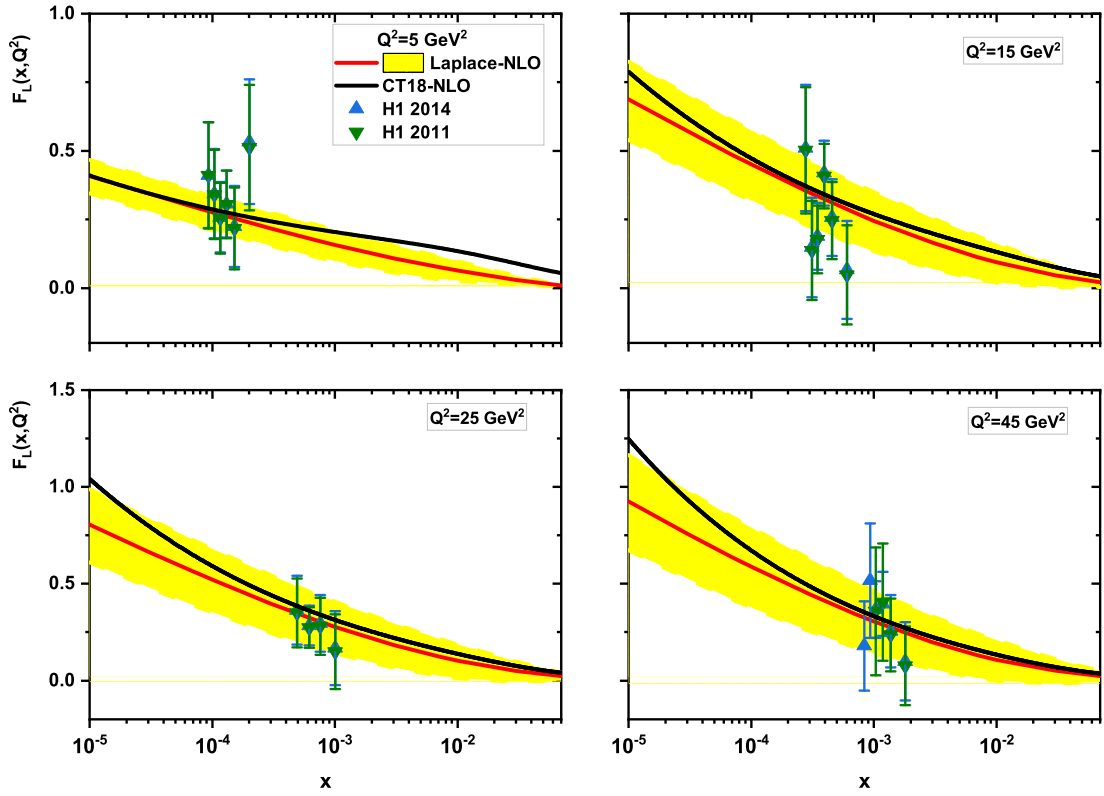}
\caption{The longitudinal structure function results at NLO
approximation with respect to the Laplace transform method
compared with the H1 experimental data (up and down triangles) [2]
as accompanied with total errors and with the CT18 NLO [24]
parametrization model. The error bands are due to the charm-quark
mass uncertainty  and the statistical errors in the
parametrization of $F_{2}(x,Q^{2})$ and its
derivative.}\label{Fig1}
\end{figure}
\begin{figure}
\includegraphics[width=1\textwidth]{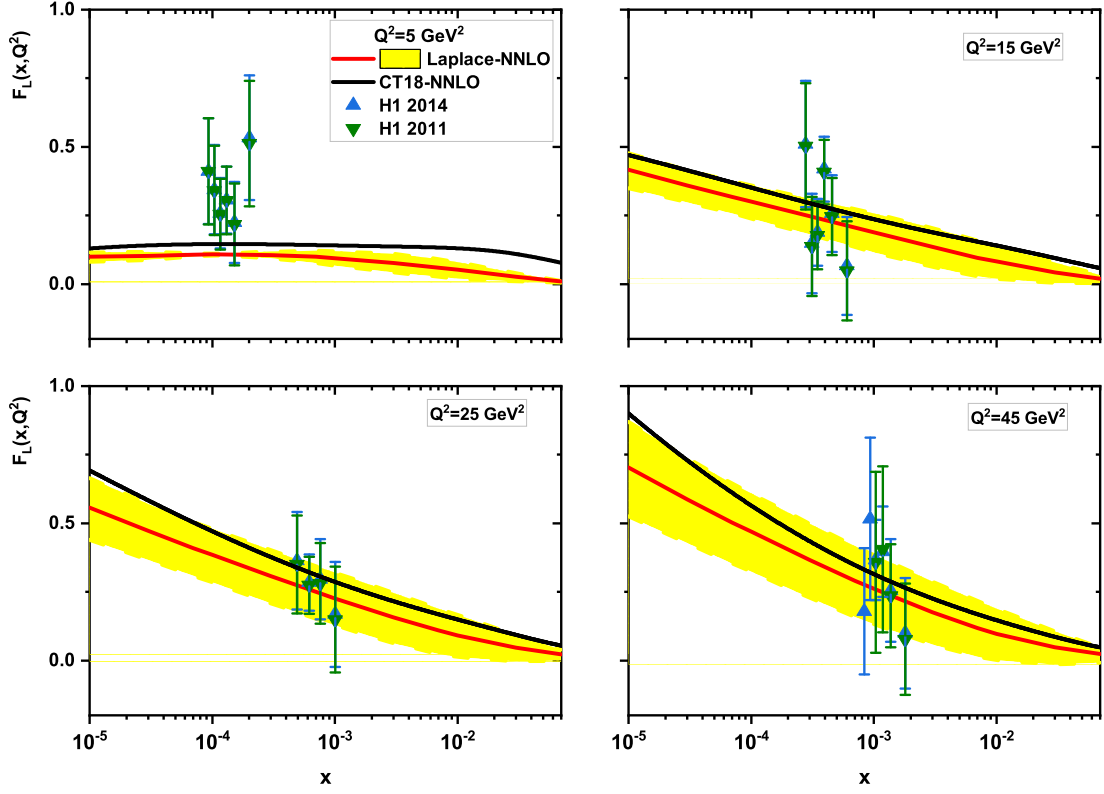}
\caption{The longitudinal structure functions at NNLO
approximation with respect to the Laplace transform method
extracted in comparison with the H1 experimental data (up and down
triangles) [2] as accompanied with total errors. The error bands
are due to the charm-quark mass uncertainty  and the statistical
errors in the parametrization of $F_{2}(x,Q^{2})$ and its
derivative. The results compared with the CT18 NNLO [24]
parametrization model.}\label{Fig2}
\end{figure}
\begin{figure}
\includegraphics[width=1\textwidth]{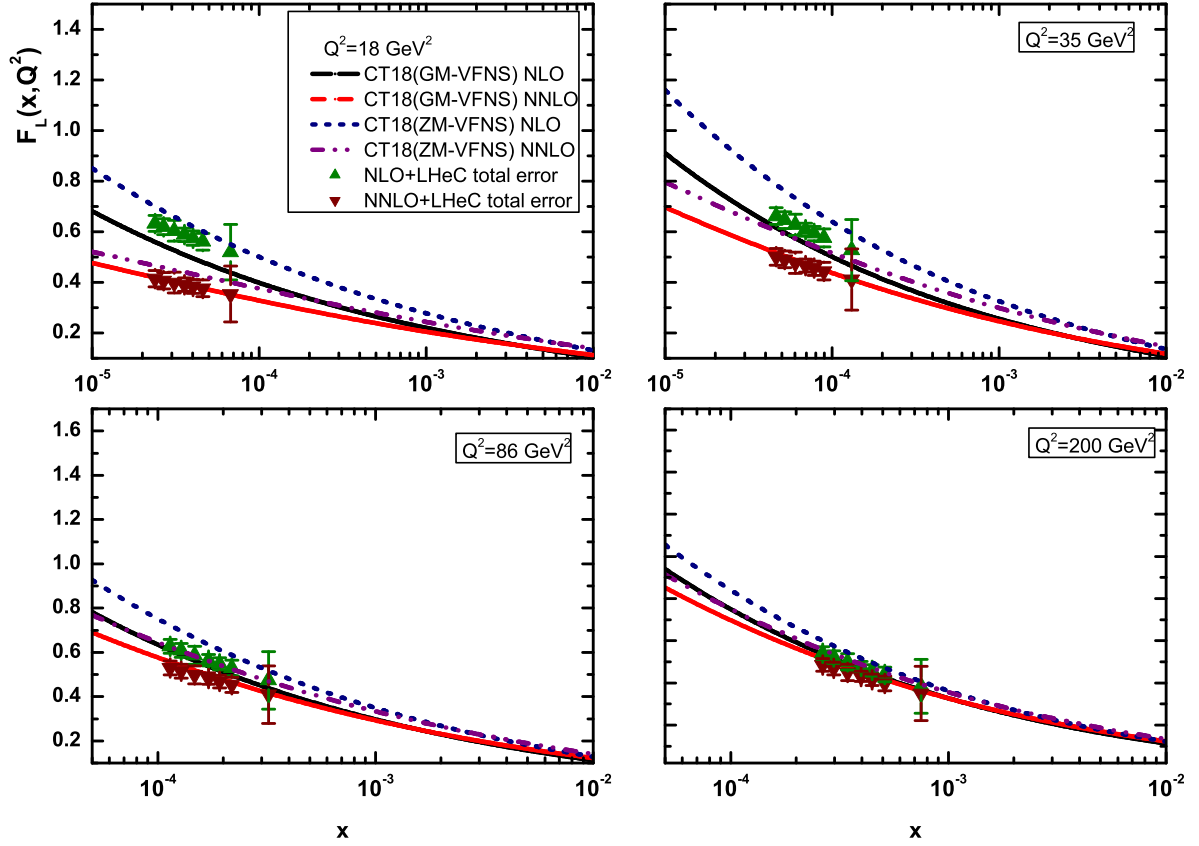}
\caption{The longitudinal structure function $F_{L}(x,Q^{2})$ with
respect to the LHeC simulated errors [9] in comparison with the
results of CT18 NLO and NNLO models [24] in the GM-VFNS and
ZM-VFNS at $Q^{2}$ values 18, 32, 86 and 200
$\mathrm{GeV}^{2}$.}\label{Fig43}
\end{figure}
\begin{figure}
\includegraphics[width=1\textwidth]{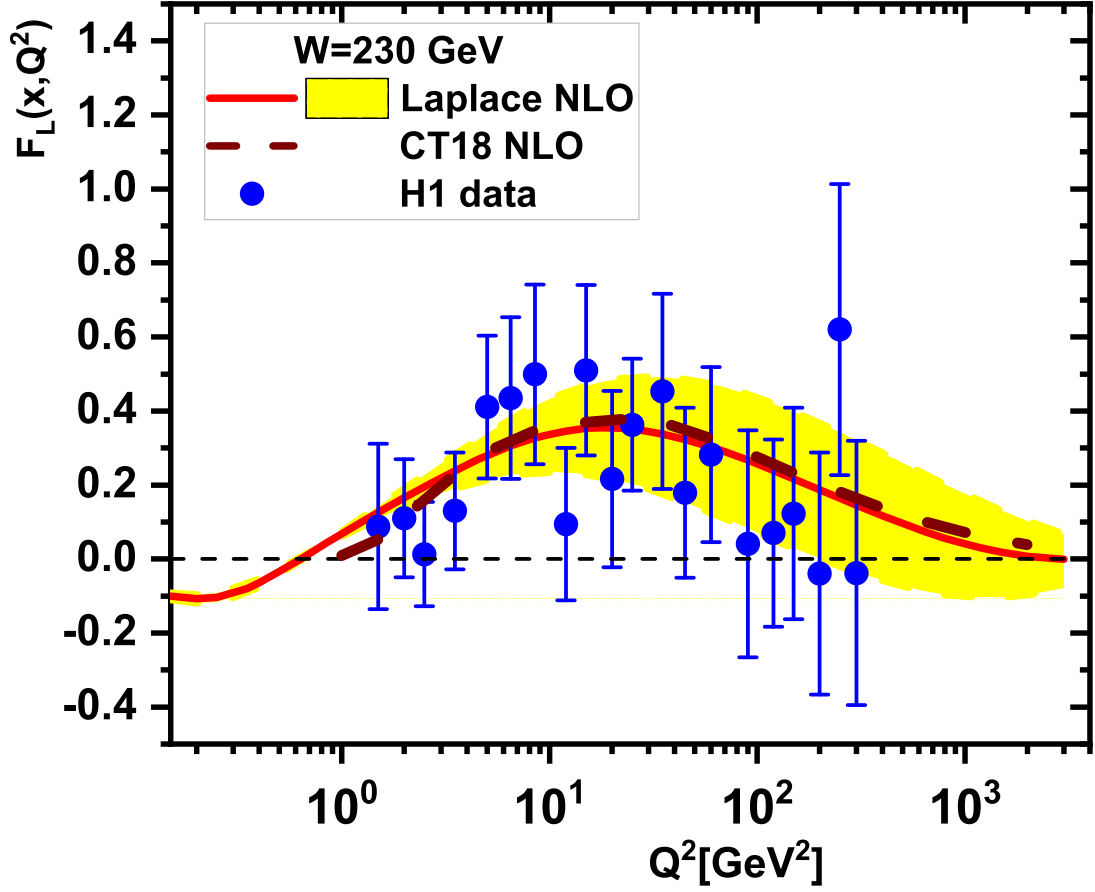}
\caption{ The extracted longitudinal structure function
$F_{L}(x,Q^{2})$ from the parametrization of $F_{2}(x,Q^{2})$ at
fixed value of the invariant mass $W=230~ \mathrm{GeV}$ (solid
curve) compared with the CT18 model [24](dashed curve) in the NLO
approximation. The error bands are due to the charm-quark mass
uncertainty  and the statistical errors in the parametrization of
$F_{2}(x,Q^{2})$ and its derivative. Experimental data are from
the H1-Collaboration, Refs.[1,2] as accompanied with total
errors.}\label{Fig4}
\end{figure}
\begin{figure}
\includegraphics[width=1\textwidth]{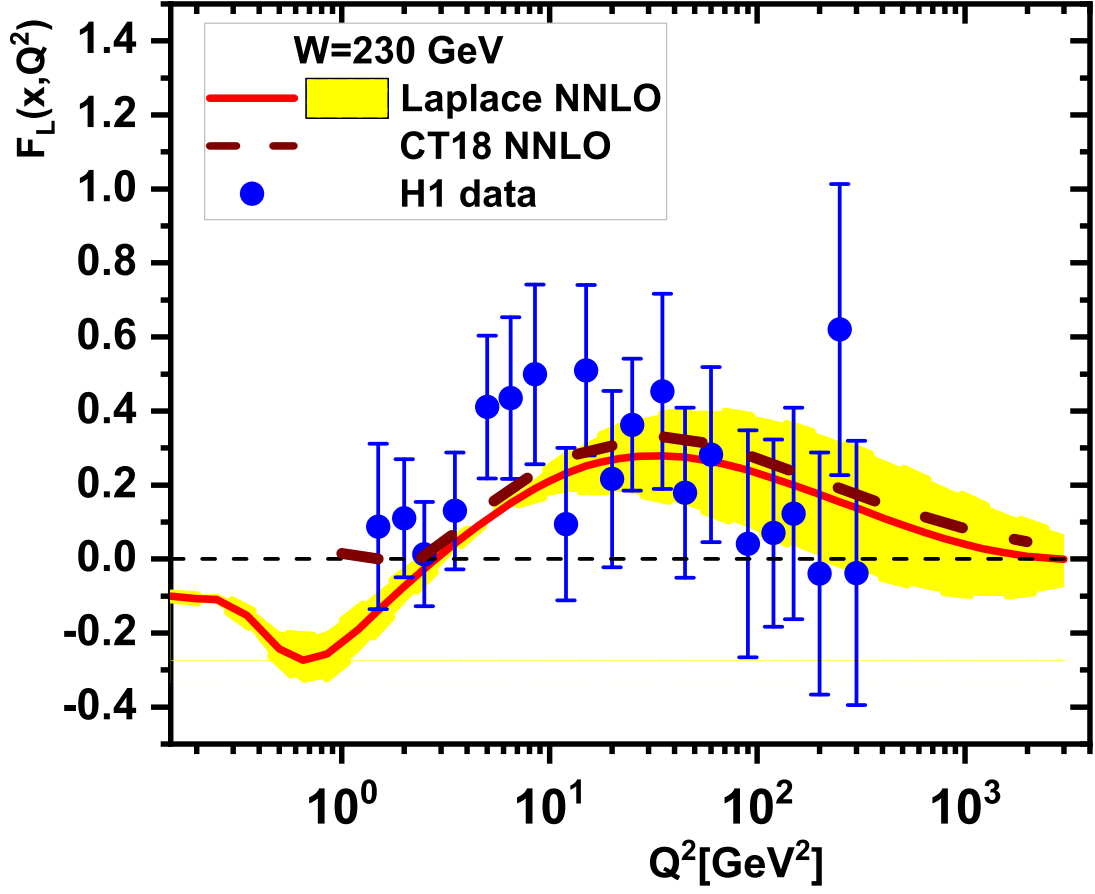}
\caption{The same as Fig.4 for the longitudinal structure function
in the NNLO approximation at $W=230~ \mathrm{GeV}$.}\label{Fig5}
\end{figure}
\begin{figure}
\includegraphics[width=1\textwidth]{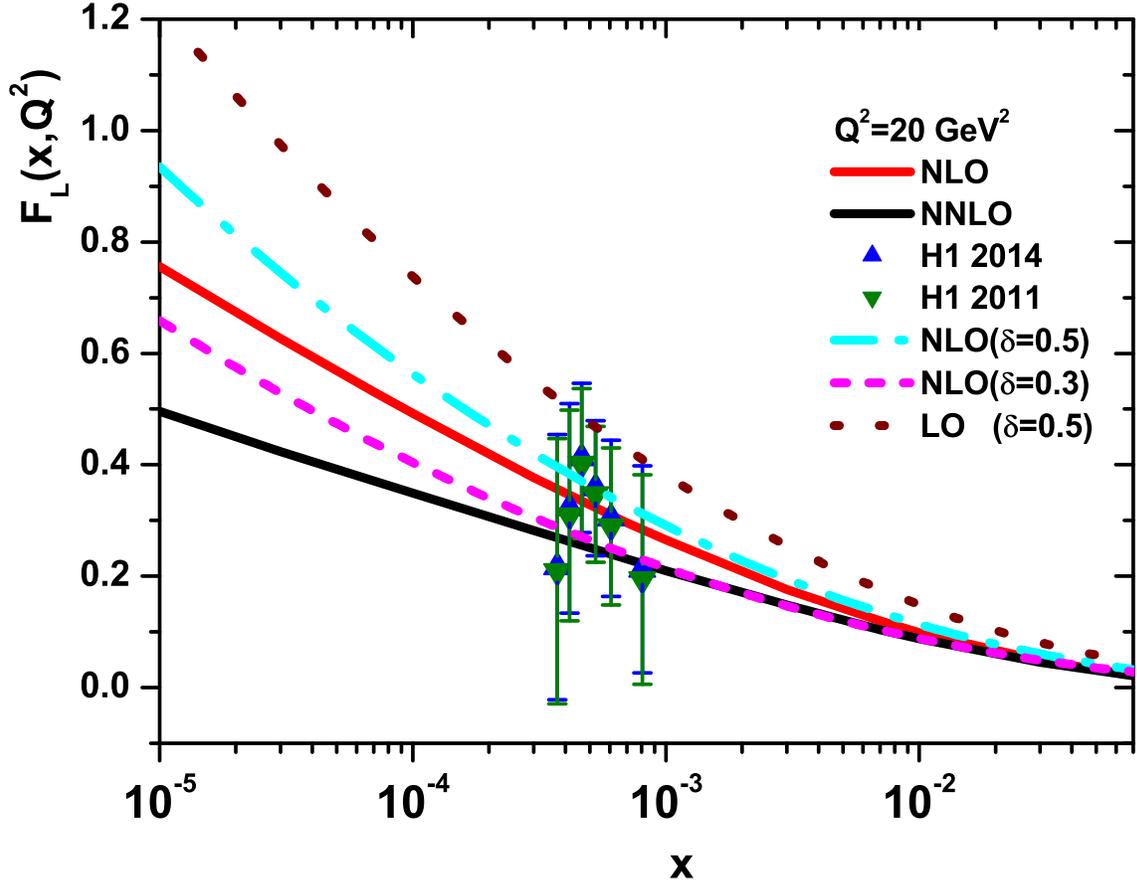}
\caption{The longitudinal structure functions at NLO and NNLO
approximations (solid curves), by the Laplace transform method,
and by the Regge-like behavior [26] at LO and NLO approximations
(Dashed and dot curves) for $\delta=0.5$ and $0.3$ at $Q^{2}=20~
\mathrm{GeV}^{2}$ compared with the H1 Collaboration data are
taken from Refs.[1,2] as accompanied with total
errors.}\label{Fig6}
\end{figure}
\begin{figure}
\includegraphics[width=1\textwidth]{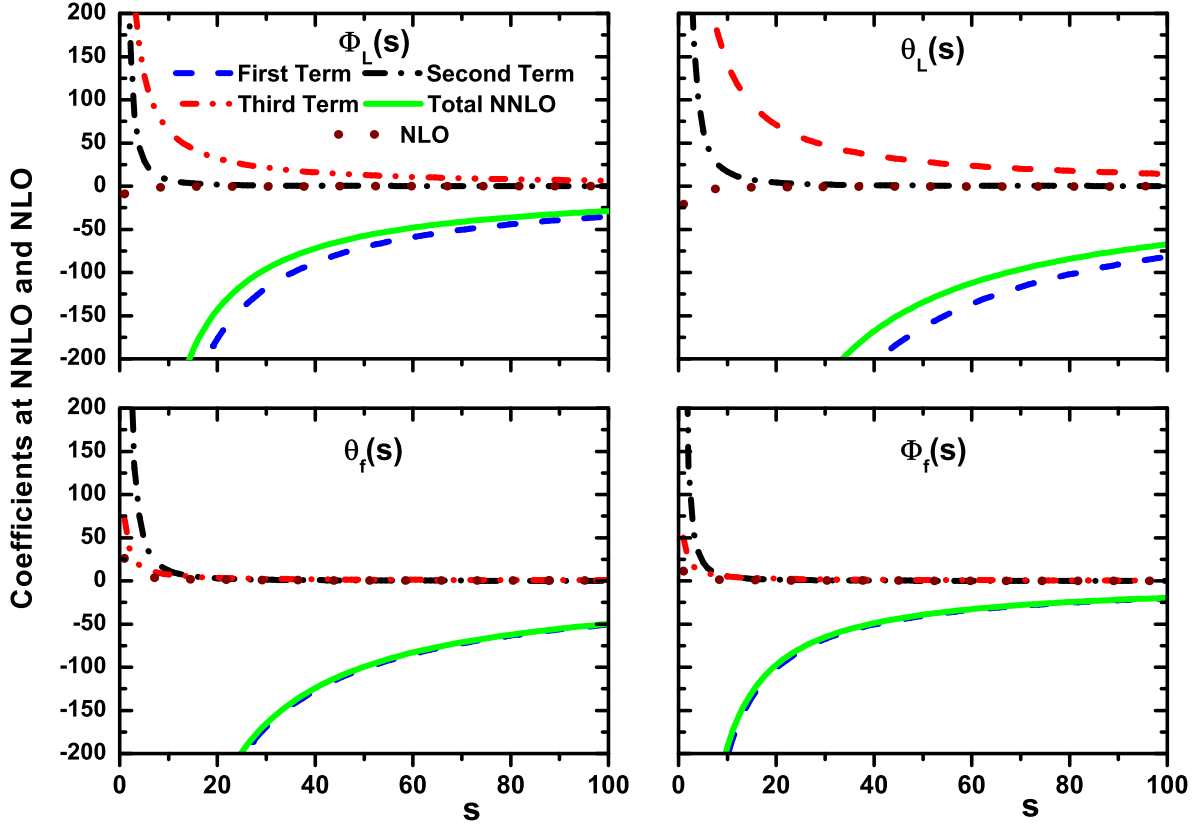}
\caption{The coefficient functions (i.e., Eqs.(14) and (15))
plotted  at small $x$  at NNLO and NLO approximations in
$s$-space. The first, second and third terms at NNLO approximation
are shown, also the total approximation coefficients at NLO
(dot-curve) compared with NNLO (solid curve) in
$s$-space.}\label{Fig7}
\end{figure}

\end{document}